\documentclass[twocolumn,aps,showpacs,showkeys,floatfix]{revtex4}
\newcommand{\bfm}[1]{\mbox{\boldmath${#1}$}}
\begin{document}
\title{Cole--Hopf Like Transformation for a Class of Coupled
 Nonlinear Schr\"odinger Equations}
\author{G. Kaniadakis}\email{kaniadakis@polito.it}
\author{E. Miraldi}\email{miraldi@polito.it}
\author{A.M. Scarfone}\email{scarfone@polito.it}
\affiliation{Dipartimento di Fisica - Politecnico di Torino \\
Corso Duca degli Abruzzi 24, 10129 Torino, Italy; \\ Istituto
Nazionale di Fisica della Materia - Unit\'a del Politecnico di
Torino}
\date{\today}
\begin {abstract}
In this paper we consider a class of coupled nonlinear
Schr\"odinger equations for the fields $\psi_{_i}$ containing
complex nonlinearities, that has been obtained by requiring that
the norms $|\psi_{_i}|^2$ are conserved densities . For this class
of equations we introduce a Cole-Hopf like transformation, whose
effect is to reduce the complex nonlinearities into real ones,
this way reducing the continuity equations of the system to the
standard bilinear form. Some examples are presented to illustrate
the applicability of the method.\\
\end {abstract}

\maketitle

Systems of coupled nonlinear Schr\"odinger equations (CNLSEs) have
attracted the attention of nonlinear physicists in recent years.
These systems emerge from the study of important phenomena
occuring in condensed matter physics, like, for instance, the
propagation of light pulse in an optical fiber \cite{Hasegawa},
propagation in a nonlinear birefringent medium
\cite{Zakharov,Ablowitz,Agrawal,Ryskin}, or in the study of
multi-species and spinor Bose-Einstein condensation \cite{Bose}.
The CNLSEs are more difficult to solve than the respective
one-component NLSEs and, generally, these systems of equations
cannot be integrated with usual methods. In the present work, by
means of a Cole-Hopf like transformation on a class of CNLSEs with
complex nonlinearities, it is possible to reduce such system of
equations to another one containing only real nonlinearities. The
method here proposed can be considered as a natural continuation
of the work presented previously in Refs. \cite{noi1,noi2}.

Let us introduce the following system of $q$ coupled nonlinear
Schr\"odinger equations in $(n+1)$-dimensions:
\begin{eqnarray}
i\,\frac{\partial\,{\bfm\psi}}{\partial\,t}+\widehat{A}\,
\Delta{\bfm\psi}+\widehat
{B}[{\bfm\psi}^\dag,\,{\bfm\psi}]\,{\bfm\psi}=0 \ ,\label{c1nlse}
\end{eqnarray}
being ${\bfm\psi}$ a $q$-component vector. Without loss of
generality, we can rearrange the system in order to reduce the
$q\times q$ matrices $\widehat A$ and $\widehat B$ to diagonal
form. The matrix $\widehat A$ has constant and real entries while
the matrix $\widehat B$ has entries which are functional dependent
on the fields $\bfm\psi$ and ${\bfm\psi}^\dag$. We use the
notation $\widehat{M}[{\bfm v}]$ to indicate the functional
dependence on the component of the vector ${\bfm
v}^t=(v_{_1},\,\cdots,\,v_{_q})$ and their spatial derivatives. In
the following, $M_{_k}$ are the $k$-th diagonal element of the
matrix $\widehat M$. Besides the complex fields $\psi_{_k}$ we
introduce the hydrodynamic real fields $\rho_{_k}$ and $S_{_k}$
defined through:
\begin{eqnarray}
\psi_{_k}({\bfm x},\,t)=\rho_{_k}^{1/2}({\bfm
x},\,t)\,\exp\left[i\,S_{_k}({\bfm x},\,t)\right] \ .
\end{eqnarray}
The matrix $\widehat B$ can be decomposed into its real $\widehat
W$ and imaginary $\widehat{\cal W}$ parts, so that Eq.
(\ref{c1nlse}) can be written as:
\begin{eqnarray}
i\,\frac{\partial\,{\bfm\psi}}{\partial\,t}+\widehat{A}\,
\Delta{\bfm\psi}+\left(\widehat
{W}[{\bfm\rho},\,{\bfm S}]+i\,\widehat{\cal W}[{\bfm\rho},\,{\bfm
S }]\right)\,{\bfm\psi}=0 \ .\label{cnlse}
\end{eqnarray}
The conservation of the quantities:
\begin{eqnarray}
N_{_k}=\int\rho_{_k}\,d^nx \ ,\label{co}
\end{eqnarray}
for $k=1,\cdots,q$ imposes the existence of $q$ continuity
equations:
\begin{eqnarray}
\frac{\partial\,\rho_{_k}}{\partial\,t}+{\bfm\nabla}\cdot{\bfm
j}_{_k}=0 \ ,\label{cont}
\end{eqnarray}
with currents ${\bfm j}_{_k}$ given by:
\begin{eqnarray}
{\bfm
j}_{_k}=2\,\left(A_{_k}\,\rho_{_k}\,{\bfm\nabla}\,S_{_k}+{\bfm
F}_{_k}\right) \ ,\label{current}
\end{eqnarray}
where the vectors ${\bfm F}_{_k}$ are related with the entries
${\cal W}_{_k}$ of the matrix $\widehat{\cal W}$ trough:
\begin{eqnarray}
{\cal W}_{_k}={1\over\rho_{_k}}\,{\bfm\nabla}\cdot{\bfm F}_{_k} \
.\label{for}
\end{eqnarray}
Note please, the structure of the elements ${\cal W}_{_k}$ in Eq.
(\ref{for}) allows the existence of the $q$ "motion constants"
defined in Eq. (\ref{co}).

Let us now introduce the transformation:
\begin{eqnarray}
{\bfm\psi}\rightarrow{\bfm\phi}=\widehat{\cal U}\,{\bfm\psi} \
,\label{tr}
\end{eqnarray}
where $\widehat{\cal U}$ is a diagonal and unitary matrix: $
\widehat{\cal U}^\dag=\widehat{\cal U}^{-1}$; with entries:
\begin{eqnarray}
{\cal U}_{_k}[{\bfm\rho},\,{\bfm
S}]=\exp\left(i\,\sigma_{_k}[{\bfm\rho},\,{\bfm S}]\right) \
,\label{u1}
\end{eqnarray}
since the generators $\sigma_{_k}[{\bfm\rho},\,{\bfm S}]$ are real
functionals. The transformation $\widehat{\cal U}$ is chosen in
order to eliminate the imaginary part of the nonlinearity in the
evolution equation for the field $\bfm \phi$, given now by:
\begin{eqnarray}
\phi_k({\bfm x},\,t)=\rho_{_k}^{1/2}({\bfm
x},\,t)\,\exp\left[i\,{\cal S}_k({\bfm x},\,t)\right] \ .
\end{eqnarray}
Since the matrix $\widehat{\cal U}$ is unitary, it results:
\begin{eqnarray}
\rho_{_k}={\bfm\psi}^\ast_{_k}\,{\bfm\psi}_{_k}=
{\bfm\phi}^\ast_{_k}\,{\bfm\phi}_{_k}
\ ,
\end{eqnarray}
while the phase ${\cal S}_{_k}$ of $\phi_{_k}$ are given by:
\begin{eqnarray}
{\cal S}_{_k}=S_{_k}+\sigma_{_k}[{\bfm\rho},\,{\bfm S}] \
.\label{phase}
\end{eqnarray}
When Eq. (\ref{phase}) is invertible, we can express the "old"
phases $S_{_k}$ as functionals of the fields $\bfm\rho$ and
$\bfm{\cal S}:\,\,S_{_k}=S_{_k}[{\bfm\rho},\,\bfm{\cal S}]$.

The generators $\sigma_{_k}$ of the transformation are defined as:
\begin{eqnarray}
{\bfm\nabla}\,\sigma_{_k}[{\bfm\rho},\,{\bfm S}]=\frac{{\bfm
F}_{_k}[{\bfm\rho},\,{\bfm S}]}{A_{_k}\,\rho_{_k}} \ ,\label{gen}
\end{eqnarray}
and from the definition (\ref{gen}) it follows the condition on
the form of the functional ${\bfm F}_{_k}$:
\begin{eqnarray}
{\bfm\nabla}\times\left(\frac{{\bfm F}_{_k}}{\rho_{_k}}\right)=0 \
.\label{condition}
\end{eqnarray}
Eq. (\ref{condition}) states the feasibility of the transformation
in any $n>1$ spatial dimension.\\ If we introduce the functional
${\bfm G}_{_k}[{\bfm\psi}^\dag,\,{\bfm\psi}]$ trough
\begin{eqnarray}
{\bfm
G}_{_\kappa}={\bfm\nabla}\,\log\,\psi_{_\kappa}+\frac{i\,{\bfm
F}_{_k}}{A_{_k}\,\rho_{_k}} \ ,\label{g}
\end{eqnarray}
the transformation (\ref{tr}), (\ref{u1}) assumes the form:
\begin{eqnarray}
{\bfm\nabla}\,\log\,\phi_{_\kappa}={\bfm
G}_{_\kappa}[{\bfm\psi}^\dag,\,{\bfm\psi}] \ .\label{16}
\end{eqnarray}
Eq. (\ref{16}) in the $n=1$ case, and when
$G_{_\kappa}[{\bfm\psi}^\dag,\,{\bfm\psi}]=\psi_{_\kappa}$
becomes:
\begin{eqnarray}
\frac{d}{d\,x}\,\log\,\phi_{_\kappa}=\psi_{_\kappa} \ ,
\end{eqnarray}
which is a Cole-Hopf transformations \cite{Ablowitz1}. In general,
when ${\bfm G}_{_\kappa}[{\bfm\psi}^\dag,\,{\bfm\psi}]$ is an
arbitrary functional, Eq. (\ref{tr}) defines a generalized
Cole-Hopf transformation.\\
 Performing now the transformation (\ref{tr}) in Eq.
(\ref{cnlse}) we obtain the following new system of CNLSEs for the
field $\bfm\phi$:
\begin{eqnarray}
i\,\frac{\partial\,{\bfm\phi}}{\partial\,t}+\widehat{A}\,
\Delta{\bfm\phi}
+\widehat{R}\,[{\bfm\rho},\,{\bfm{\cal S}}]\,{\bfm\phi}=0 \
,\label{cnlse2}
\end{eqnarray}
which contains a purely real nonlinearity
$\widehat{R}\,[{\bfm\rho},\,{\bfm{\cal S}}]$ with entries given
by:
\begin{eqnarray}
R_{_k}\,[{\bfm\rho},\,{\bfm{\cal
S}}]=W_{_k}-A_{_k}\,({\bfm\nabla}\,\sigma_{_k})^2+\frac{{\bfm
J}_{_k}\cdot{\bfm\nabla}\,\sigma_{_k}}{\rho_{_k}}+
\frac{\partial\,\sigma_{_k}}{\partial\,t}
\ .
\end{eqnarray}
Note that $\widehat R$ depends implicitly on the fields $\bfm{\cal
S}$, since Eq. (\ref{phase}) defines $\bfm S$ as a function of
$\bfm\rho$ and $\bfm{\cal S}$. It is easy now to verify that Eq.
(\ref{cnlse2}) admits the following continuity equations:
\begin{eqnarray}
\frac{\partial\,\rho_{_k}}{\partial\,t}+{\bfm\nabla}\cdot{\bfm
J}_k=0 \ ,
\end{eqnarray}
where the currents ${\bfm J}_{_k}$ assumes the standard form of
the linear quantum mechanics and are given by:
\begin{eqnarray}
{\bfm J}_{_k}=2\,A_{_k}\,\rho_{_k}\,{\bfm\nabla}\,{\cal S}_{_k} \
.\label{cor}
\end{eqnarray}

In the following, in order to show the feasibility of the proposed
method, we consider some CNLSE, already studied in literature by
different authors.\\

We start by considering the CNLSEs \cite{Coskun,Kivshar} with
complex nonlinearity given by:
\begin{eqnarray}
\widehat{W}_{_k}[{\bfm\rho},\,{\bfm S}]
&=&{\bfm\delta}_{_k}\cdot{\bfm\nabla}\,S_{_k}-
\gamma_{_k}\,\rho_{_k}-2\sum_{ j\not=k}\gamma_{_j}\,\rho_{_j} \
,\\ \widehat{\cal W}_{_k}[{\bfm\rho}]&=&-\frac{1}{2}\,{\bfm
\delta}_{_k}\,\cdot{\bfm\nabla}\,\log\,\rho_{_k} \ .\label{u2}
\end{eqnarray}
It is easy to verify that the transformation generated by:
\begin{eqnarray}
\sigma_{_k}=-{1\over2}\,{{\bfm\delta}_{_k}\over A_{_k}}\cdot{\bfm
x} \ ,
\end{eqnarray}
eliminates the imaginary part in the nonlinearity and changes the
real part in:
\begin{eqnarray}
\widehat{R}_{_k}[{\bfm\rho}]=-\gamma_{_k}\,\rho_{_k}-2\sum_{
j\not=k}\gamma_{_j}\,\rho_{_j}-{1\over4}
\frac{{\bfm\delta}_{_k}^2}{A_{_k}}
\ ,
\end{eqnarray}
so that the new CNLSEs become a system of coupled cubic nonlinear
Schr\"odinger equations.

We consider now the 1-dimensional class of CNLSEs with complex
nonlinearity given by:
\begin{eqnarray}
\nonumber W_{_k}[{\bfm\rho},\,{\bfm S}]&=&\sum_j\,
\rho_{_j}\left(\beta_{_{kj}}\frac{\partial\,S_{_k}}{\partial\,x}
+\gamma_{_{kj}}\,
\frac{\partial\,S_{_j}}{\partial\,x}\right)\\&+&\sum_{j,i}
\lambda_{_{kji}}\,\rho_{_j}\,\rho_{_i} \  ,\\
\nonumber {\cal W}_{_k}[{\bfm\rho}]&=&2\,\delta_{_{kk}}\,
\frac{\partial\,\rho_{_k}}{\partial\,x}+\sum_{j\not=k}\delta_{_{kj}}
\left(\frac{\partial\,\rho_{_j}}{\partial\,x}
+\frac{\rho_{_j}}{\rho_{_k}}\,\frac{\partial\,\rho_{_k}}
{\partial\,x}\right)\ .\\
\end{eqnarray}
Remark that this class contains, as particular cases some very
important equations, well known in literature. For instance, in
the case $q=1$ we obtain, for $\delta=\lambda=0$, the Jackiw
equation \cite{Jackiw}, for $4\,\delta+\beta+\gamma=0$ and
$\lambda=0$ the Chen-Lee-Liu equation \cite{Chen} and for
$4\,\delta+3\,(\beta+\gamma)=0$ and $\lambda=0$ the Kaup-Newell
equation \cite{Kaup}. In the case $q=2$ with $\lambda_{_{kij}}=0$,
the above CNLSEs have been studied recently in Ref.
\cite{Wadati,Wadati1,Tsuchida}).

It is easy to verify that the expression of ${\cal W}_{_k}$ allows
the conservation of the quantities $\int\rho_{_k}\,d^nx$ and, in
this case, the current appearing in the continuity equations
(\ref{cont}) takes the form:
\begin{eqnarray}
j_{_k}=2\,\rho_{_k}\,\left[A_{_k}\,\frac{\partial\,S_{_k}}
{\partial\,x}+\sum_j\delta_{_{kj}}\,\rho_{_j}\right] \ .
\end{eqnarray}
The transformation which eliminates the imaginary part of the
nonlinearity has generator $\sigma_{_k}$ given by:
\begin{eqnarray}
\sigma_{_k}={1\over A_{_k}}\sum_j\delta_{_{kj}}\int^x\rho_{_j}\,dx
\ ,
\end{eqnarray}
while the real nonlinearity in the new system of CNLSEs for the
fields ${\bfm\phi}$ becomes:
\begin{eqnarray}
\nonumber R_{_k}[{\bfm\rho},\,{\bfm
S}]&=&\sum_j\left[\left(\beta_{_{kj}}+2\,\delta_{_{kj}}\right)
\,\frac{\partial\,{\cal
S}_{_k}}{\partial\,x}\right.\\
&+&\left.\left(\gamma_{_{kj}}-2\,\frac{A_{_j}}
{A_{_k}}\,\delta_{_{kj}}\right)\,
\frac{\partial\,{\cal S}_{_j}}{\partial\,x}\right]\,\rho_{_j}\\
\nonumber &-&\sum_{j,i}\left[\frac{\delta_{_{kj}}}{A_{_k}}\,
\left(\delta_{_{ki}}+\beta_{_{ki}}
\right)+\frac{\gamma_{_{kj}}}{A_{_j}}\,\delta_{_{ji}}-
\lambda_{_{kji}} \right]\,\rho_{_j}\,\rho_{_i} \ .
\end{eqnarray}
Finally we summarize some particular cases of these new CNLSEs:
\begin{enumerate}
\item For
$\beta_{_{kj}}=-2\,\delta_{_{kj}},\,\,\gamma_{_{kj}}=2\,A_{_j}\,
\delta_{_{kj}}/A_{_k},\,\,
\lambda_{_{kji}}=\delta_{_{kj}}\left(2\,\delta_{_{ji}}-
\delta_{_{ki}}\right)/A_{_k},$
we obtain a system of decoupled linear Sch\"odringer equations:
\begin{eqnarray}
i\,\frac{\partial\,\phi_{_k}}{\partial\,t}+A_{_k}\,
\frac{\partial^2\,\phi_{_k}}{\partial\,x^2}=0
\ ,
\end{eqnarray}
\item For $\beta_{_{kj}}=-2\,\delta_{_{kj}}$ with $k\not=j,\,\,
\gamma_{_{kj}}=2\,A_{_j}\,\delta_{_{kj}}/A_{_k},\,$
\begin{eqnarray}
\left\{
\begin{array}{l}
\lambda_{_{kkk}}=\delta_{_{kk}}\,\left(\beta_{_{kk}}+3\,
\delta_{_{kk}}\right)/A_{_k}
\ ,\\
\lambda_{_{kjk}}=\delta_{_{kj}}\left(\beta_{_{kk}}+
\delta_{_{kk}}+2\,\delta_{_{jk}}\right)/A_{_k}
\ ,\\
\lambda_{_{kki}}=\delta_{_{kk}}\,\delta_{_{ki}}/A_{_k} \ ,\\
\lambda_{_{kji}}=\delta_{_{kj}}\left(2\,\delta_{_{ji}}-
\delta_{_{ki}}\right)/A_{_k}, \scriptstyle{{\rm for} k\not=j\not=i
{\rm and} \kappa\not=j=i} \ ,
\end{array}
\right.
\end{eqnarray}
we obtain the following system of decoupled Jackiw-like NLSEs:
\begin{eqnarray}
i\,\frac{\partial\,\phi_{_k}}{\partial\,t}+A_{_k}\,
\frac{\partial^2\,\phi_{_k}}{\partial\,x^2}
+\eta_{_k}\,J_{_k}\,\phi_{_k}=0 \ ,\label{ex1}
\end{eqnarray}
with $\eta_{_k}=(\beta_{_{kk}}+2\,\delta_{_{kk}})/(2\,A_{_k})$ and
$J_{_k}$ given by Eq. (\ref{cor}).
\item Finally, for
$\beta_{_{kj}}=-2\,\delta_{_{kj}},\,\,\lambda_{_{kji}}=
\gamma_{_{kj}}\,\delta_{_{ji}}/A_{_j}-
\delta_{_{kj}}\,\delta_{_{ki}}/A_{_k}$ we obtain the CNLSEs:
\begin{eqnarray}
i\,\frac{\partial\,\phi_{_k}}{\partial\,t}+A_{_k}\,
\frac{\partial^2\,\phi_{_k}}{\partial\,x^2}
+\sum_j\eta_{_{kj}}\,J_{_j}\,\phi_{_k}=0 \ ,\label{34}
\end{eqnarray}
being
$\eta_{_{kj}}=(\gamma_{_{kj}}-2\,A_{_j}\,
\delta_{_{kj}}/A_{_k})/(2\,A_{_j})$.
The nonlinear term in Eq. (\ref{34}) has been considered in Ref.
\cite{Calogero}.
\end{enumerate}

In conclusion, we have generalized the method introduced in Ref.
\cite{noi1,noi2} to the case of CNLSEs with complex
nonlinearities. In order to assure the conservation of the
particle number $N_{_i}\,\,(i=1\cdots q)$, for each species, we
assumed the system admits $q$ continuity equations. The condition
given by Eq. (\ref{condition}) must be satisfied if the spatial
dimension is $n>1$. Differently from \cite{noi1,noi2}, in the
present approach we do not require the existence of a Lagrangian
for the system. In particular, for $q=1$, the present theory can
be viewed as a generalization to the case of noncanonical systems
of the canonical theory of Ref. \cite{noi1}.

The general transformation here introduced gives us the
possibility to deal, in the frame of a unifying scheme, different
coupled nonlinear equations, some of which already known in
literature.

\vfill\eject
\end{document}